\documentclass[aps,prd,twocolumn,amsmath,amssymb,amsfonts,nofootinbib,superscriptaddress,altaffilletter]{revtex4-1}
\usepackage{array}
\newcolumntype{P}[1]{>{\centering\arraybackslash}p{#1}}

\usepackage{graphicx}
\usepackage{amssymb}
\usepackage{amsmath}
\usepackage{longtable}
\usepackage{color}
\usepackage{etoolbox}
\usepackage{supertabular}
\usepackage{dcolumn}
\usepackage{ wasysym }

\def\CR{\textrm{CR}}

\def\Tobs{T_{\textrm{\mbox{\tiny{obs}}}}}
\def\Tcoh{T_{\textrm{\mbox{\tiny{coh}}}}}
\def\Tref{T_{\textrm{\mbox{\tiny{ref}}}}}

\def\EatH{Einstein@Home }

\def\sci#1#2{#1\times10^{#2}}

\newcommand{\prior}[1]{\MakeLowercase{#1}}

\newcommand{\avgX}[1]{\left\langle #1 \right\rangle}	
\newcommand{\avgSeg}[1]{\overline{#1}}			

\newcommand{\cF}{c_*}

\newcommand{\Gauss}{\mathrm{\MakeUppercase{G}}}
\newcommand{\Signal}{{\mathrm{\MakeUppercase{S}}}}
\newcommand{\Line}{{\mathrm{\MakeUppercase{L}}}}
\newcommand{\Noise}{{\Gauss\Line}}

\providecommand{\sc}[1]{\widehat{#1}}
\renewcommand{\sc}[1]{\widehat{#1}}

\newcommand{\OSNsc}{\sc{O}_{{\Signal\Noise}}}	
\newcommand{\OSLsc}{\sc{O}_{{\Signal\Line}}}

\newcommand{\OLGsc}{\sc{O}_{{\Line\Gauss}}}

\newcommand{\oSLsc}{\prior{\OSLsc}}
\newcommand{\oLGsc}{\prior{\OLGsc}}

\newcommand{\rsc}{\sc{r}}

\newcommand{\lineprobsc}{\sc{p}_{{\Line}}}

\newcommand{\F}{\mathcal{F}}		

\newcommand{\Fth}{\F_*}
\newcommand{\Ftho}{\Fth^{(0)}}

\newcommand{\scF}{\sc{\F}}
\newcommand{\scFX}{\scF^X}

\newcommand{\scFmaxG}{\scF''_{\mathrm{max}}}
\newcommand{\scFth}{\scF_*}
\newcommand{\scFtho}{\scF_*^{(0)}}
\newcommand{\scrX}{\sc{r}^X}

\newcommand{\avF}{\avgSeg{\F}}
\newcommand{\avFX}{\avF^X}

\newcommand{\pFA}{p_{\mathrm{FA}}}

\newcommand{\Ndet}{{N_{\mathrm{det}}}}
\newcommand{\Nseg}{{N_{\mathrm{seg}}}}

\newcommand{\eto}{\ensuremath{\mathrm{e}^}}	

\newcommand{\sft}{{\mathrm{SFT}}}

\newcommand{\Psft}{\mathcal{P}} 
\newcommand{\Psftthr}{\Psft_{\mathrm{thr}}} 
\newcommand{\Nsft}{N_\sft}

\newcommand{\paramfdotlo}{\ensuremath{-\sci{2.67}{-9}~\mathrm{Hz/s}}} 
\newcommand{\paramfdothi}{\ensuremath{\sci{3.39}{-10}~\mathrm{Hz/s}}} 
\newcommand{\paramWUcputimeHours}{6} 
\newcommand{\paramWUavgskypointsInt}{13} 
\newcommand{\paramWUtotaltemplates}{\ensuremath{\sci{4.9}{9}}} 
\newcommand{\paramtotalWUsmillions}{12.7} 
\newcommand{\paramtotaltemplates}{\ensuremath{\sci{6.3}{16}}} 
\newtoggle{fullauthorlist}
\toggletrue{fullauthorlist}

\newtoggle{endauthorlist}
\toggletrue{endauthorlist}

\begin{document}

\title{ 
Results of the deepest all-sky survey for continuous gravitational waves on LIGO S6 data running on the \EatH volunteer distributed computing project
}


\iftoggle{endauthorlist}{
  %
  %
  \let\mymaketitle\maketitle
  \let\myauthor\author
  \let\myaffiliation\affiliation
  \author{The LIGO Scientific Collaboration}
  \author{The Virgo Collaboration}
}{
  %
  %
  \iftoggle{fullauthorlist}{
    \author{%
B.~P.~Abbott,$^{1}$  
R.~Abbott,$^{1}$  
T.~D.~Abbott,$^{2}$  
M.~R.~Abernathy,$^{3}$  
F.~Acernese,$^{4,5}$ 
K.~Ackley,$^{6}$  
C.~Adams,$^{7}$  
T.~Adams,$^{8}$ 
P.~Addesso,$^{9}$  
R.~X.~Adhikari,$^{1}$
% ...
\\
\medskip
(LIGO Scientific Collaboration and Virgo Collaboration) 
\\
\medskip
{{}$^{*}$Deceased, March 2016. {}$^{\dag}$Deceased, May 2015. {}$^{\ddag}$Deceased, March 2015. }%
}\noaffiliation
\affiliation {LIGO, California Institute of Technology, Pasadena, CA 91125, USA }
\affiliation {Louisiana State University, Baton Rouge, LA 70803, USA }
\affiliation {American University, Washington, D.C. 20016, USA }
\affiliation {Universit\`a di Salerno, Fisciano, I-84084 Salerno, Italy }
\affiliation {INFN, Sezione di Napoli, Complesso Universitario di Monte S.Angelo, I-80126 Napoli, Italy }
\affiliation {University of Florida, Gainesville, FL 32611, USA }
\affiliation {LIGO Livingston Observatory, Livingston, LA 70754, USA }
\affiliation {Laboratoire d'Annecy-le-Vieux de Physique des Particules (LAPP), Universit\'e Savoie Mont Blanc, CNRS/IN2P3, F-74941 Annecy-le-Vieux, France }
\affiliation {University of Sannio at Benevento, I-82100 Benevento, Italy and INFN, Sezione di Napoli, I-80100 Napoli, Italy }
\affiliation {Albert-Einstein-Institut, Max-Planck-Institut f\"ur Gravi\-ta\-tions\-physik, D-30167 Hannover, Germany }
\affiliation {Nikhef, Science Park, 1098 XG Amsterdam, The Netherlands }
\affiliation {LIGO, Massachusetts Institute of Technology, Cambridge, MA 02139, USA }
\affiliation {Instituto Nacional de Pesquisas Espaciais, 12227-010 S\~{a}o Jos\'{e} dos Campos, S\~{a}o Paulo, Brazil }
\affiliation {INFN, Gran Sasso Science Institute, I-67100 L'Aquila, Italy }
\affiliation {INFN, Sezione di Roma Tor Vergata, I-00133 Roma, Italy }
\affiliation {Inter-University Centre for Astronomy and Astrophysics, Pune 411007, India }
\affiliation {International Centre for Theoretical Sciences, Tata Institute of Fundamental Research, Bangalore 560012, India }
\affiliation {University of Wisconsin-Milwaukee, Milwaukee, WI 53201, USA }
\affiliation {Leibniz Universit\"at Hannover, D-30167 Hannover, Germany }
\affiliation {Universit\`a di Pisa, I-56127 Pisa, Italy }
\affiliation {INFN, Sezione di Pisa, I-56127 Pisa, Italy }
\affiliation {Australian National University, Canberra, Australian Capital Territory 0200, Australia }
\affiliation {The University of Mississippi, University, MS 38677, USA }
\affiliation {California State University Fullerton, Fullerton, CA 92831, USA }
\affiliation {LAL, Univ. Paris-Sud, CNRS/IN2P3, Universit\'e Paris-Saclay, Orsay, France }
\affiliation {Chennai Mathematical Institute, Chennai 603103, India }
\affiliation {Universit\`a di Roma Tor Vergata, I-00133 Roma, Italy }
\affiliation {University of Southampton, Southampton SO17 1BJ, United Kingdom }
\affiliation {Universit\"at Hamburg, D-22761 Hamburg, Germany }
\affiliation {INFN, Sezione di Roma, I-00185 Roma, Italy }
\affiliation {Albert-Einstein-Institut, Max-Planck-Institut f\"ur Gravitations\-physik, D-14476 Potsdam-Golm, Germany }
\affiliation {APC, AstroParticule et Cosmologie, Universit\'e Paris Diderot, CNRS/IN2P3, CEA/Irfu, Observatoire de Paris, Sorbonne Paris Cit\'e, F-75205 Paris Cedex 13, France }
\affiliation {Montana State University, Bozeman, MT 59717, USA }
\affiliation {Universit\`a di Perugia, I-06123 Perugia, Italy }
\affiliation {INFN, Sezione di Perugia, I-06123 Perugia, Italy }
\affiliation {European Gravitational Observatory (EGO), I-56021 Cascina, Pisa, Italy }
\affiliation {Syracuse University, Syracuse, NY 13244, USA }
\affiliation {SUPA, University of Glasgow, Glasgow G12 8QQ, United Kingdom }
\affiliation {LIGO Hanford Observatory, Richland, WA 99352, USA }
\affiliation {Wigner RCP, RMKI, H-1121 Budapest, Konkoly Thege Mikl\'os \'ut 29-33, Hungary }
\affiliation {Columbia University, New York, NY 10027, USA }
\affiliation {Stanford University, Stanford, CA 94305, USA }
\affiliation {Universit\`a di Padova, Dipartimento di Fisica e Astronomia, I-35131 Padova, Italy }
\affiliation {INFN, Sezione di Padova, I-35131 Padova, Italy }
\affiliation {CAMK-PAN, 00-716 Warsaw, Poland }
\affiliation {University of Birmingham, Birmingham B15 2TT, United Kingdom }
\affiliation {Universit\`a degli Studi di Genova, I-16146 Genova, Italy }
\affiliation {INFN, Sezione di Genova, I-16146 Genova, Italy }
\affiliation {RRCAT, Indore MP 452013, India }
\affiliation {Faculty of Physics, Lomonosov Moscow State University, Moscow 119991, Russia }
\affiliation {SUPA, University of the West of Scotland, Paisley PA1 2BE, United Kingdom }
\affiliation {University of Western Australia, Crawley, Western Australia 6009, Australia }
\affiliation {Department of Astrophysics/IMAPP, Radboud University Nijmegen, P.O. Box 9010, 6500 GL Nijmegen, The Netherlands }
\affiliation {Artemis, Universit\'e C\^ote d'Azur, CNRS, Observatoire C\^ote d'Azur, CS 34229, Nice cedex 4, France }
\affiliation {Institut de Physique de Rennes, CNRS, Universit\'e de Rennes 1, F-35042 Rennes, France }
\affiliation {Washington State University, Pullman, WA 99164, USA }
\affiliation {Universit\`a degli Studi di Urbino ``Carlo Bo,'' I-61029 Urbino, Italy }
\affiliation {INFN, Sezione di Firenze, I-50019 Sesto Fiorentino, Firenze, Italy }
\affiliation {University of Oregon, Eugene, OR 97403, USA }
\affiliation {Laboratoire Kastler Brossel, UPMC-Sorbonne Universit\'es, CNRS, ENS-PSL Research University, Coll\`ege de France, F-75005 Paris, France }
\affiliation {Carleton College, Northfield, MN 55057, USA }
\affiliation {Astronomical Observatory Warsaw University, 00-478 Warsaw, Poland }
\affiliation {VU University Amsterdam, 1081 HV Amsterdam, The Netherlands }
\affiliation {University of Maryland, College Park, MD 20742, USA }
\affiliation {Center for Relativistic Astrophysics and School of Physics, Georgia Institute of Technology, Atlanta, GA 30332, USA }
\affiliation {Laboratoire des Mat\'eriaux Avanc\'es (LMA), CNRS/IN2P3, F-69622 Villeurbanne, France }
\affiliation {Universit\'e Claude Bernard Lyon 1, F-69622 Villeurbanne, France }
\affiliation {Universit\`a di Napoli ``Federico II,'' Complesso Universitario di Monte S.Angelo, I-80126 Napoli, Italy }
\affiliation {NASA/Goddard Space Flight Center, Greenbelt, MD 20771, USA }
\affiliation {RESCEU, University of Tokyo, Tokyo, 113-0033, Japan. }
\affiliation {Tsinghua University, Beijing 100084, China }
\affiliation {Texas Tech University, Lubbock, TX 79409, USA }
\affiliation {The Pennsylvania State University, University Park, PA 16802, USA }
\affiliation {National Tsing Hua University, Hsinchu City, 30013 Taiwan, Republic of China }
\affiliation {Charles Sturt University, Wagga Wagga, New South Wales 2678, Australia }
\affiliation {West Virginia University, Morgantown, WV 26506, USA }
\affiliation {University of Chicago, Chicago, IL 60637, USA }
\affiliation {Caltech CaRT, Pasadena, CA 91125, USA }
\affiliation {Korea Institute of Science and Technology Information, Daejeon 305-806, Korea }
\affiliation {Universit\`a di Roma ``La Sapienza,'' I-00185 Roma, Italy }
\affiliation {University of Brussels, Brussels 1050, Belgium }
\affiliation {Sonoma State University, Rohnert Park, CA 94928, USA }
\affiliation {Center for Interdisciplinary Exploration \& Research in Astrophysics (CIERA), Northwestern University, Evanston, IL 60208, USA }
\affiliation {University of Minnesota, Minneapolis, MN 55455, USA }
\affiliation {The University of Melbourne, Parkville, Victoria 3010, Australia }
\affiliation {Institute for Plasma Research, Bhat, Gandhinagar 382428, India }
\affiliation {The University of Sheffield, Sheffield S10 2TN, United Kingdom }
\affiliation {The University of Texas Rio Grande Valley, Brownsville, TX 78520, USA }
\affiliation {Universit\`a di Trento, Dipartimento di Fisica, I-38123 Povo, Trento, Italy }
\affiliation {INFN, Trento Institute for Fundamental Physics and Applications, I-38123 Povo, Trento, Italy }
\affiliation {Cardiff University, Cardiff CF24 3AA, United Kingdom }
\affiliation {Montclair State University, Montclair, NJ 07043, USA }
\affiliation {MTA E\"otv\"os University, ``Lendulet'' Astrophysics Research Group, Budapest 1117, Hungary }
\affiliation {National Astronomical Observatory of Japan, 2-21-1 Osawa, Mitaka, Tokyo 181-8588, Japan }
\affiliation {School of Mathematics, University of Edinburgh, Edinburgh EH9 3FD, United Kingdom }
\affiliation {Indian Institute of Technology, Gandhinagar Ahmedabad Gujarat 382424, India }
\affiliation {University of Szeged, D\'om t\'er 9, Szeged 6720, Hungary }
\affiliation {Embry-Riddle Aeronautical University, Prescott, AZ 86301, USA }
\affiliation {Tata Institute of Fundamental Research, Mumbai 400005, India }
\affiliation {INAF, Osservatorio Astronomico di Capodimonte, I-80131, Napoli, Italy }
\affiliation {University of Michigan, Ann Arbor, MI 48109, USA }
\affiliation {Rochester Institute of Technology, Rochester, NY 14623, USA }
\affiliation {NCSA, University of Illinois at Urbana-Champaign, Urbana, Illinois 61801, USA }
\affiliation {Universitat de les Illes Balears, IAC3---IEEC, E-07122 Palma de Mallorca, Spain }
\affiliation {University of Bia{\l }ystok, 15-424 Bia{\l }ystok, Poland }
\affiliation {SUPA, University of Strathclyde, Glasgow G1 1XQ, United Kingdom }
\affiliation {IISER-TVM, CET Campus, Trivandrum Kerala 695016, India }
\affiliation {Canadian Institute for Theoretical Astrophysics, University of Toronto, Toronto, Ontario M5S 3H8, Canada }
\affiliation {Institute of Applied Physics, Nizhny Novgorod, 603950, Russia }
\affiliation {Pusan National University, Busan 609-735, Korea }
\affiliation {Hanyang University, Seoul 133-791, Korea }
\affiliation {University of Adelaide, Adelaide, South Australia 5005, Australia }
\affiliation {NCBJ, 05-400 \'Swierk-Otwock, Poland }
\affiliation {IM-PAN, 00-956 Warsaw, Poland }
\affiliation {Monash University, Victoria 3800, Australia }
\affiliation {Seoul National University, Seoul 151-742, Korea }
\affiliation {The Chinese University of Hong Kong, Shatin, NT, Hong Kong SAR, China }
\affiliation {University of Alabama in Huntsville, Huntsville, AL 35899, USA }
\affiliation {University of Massachusetts-Amherst, Amherst, MA 01003, USA }
\affiliation {ESPCI, CNRS, F-75005 Paris, France }
\affiliation {Universit\`a di Camerino, Dipartimento di Fisica, I-62032 Camerino, Italy }
\affiliation {Southern University and A\&M College, Baton Rouge, LA 70813, USA }
\affiliation {College of William and Mary, Williamsburg, VA 23187, USA }
\affiliation {Instituto de F\'\i sica Te\'orica, University Estadual Paulista/ICTP South American Institute for Fundamental Research, S\~ao Paulo SP 01140-070, Brazil }
\affiliation {University of Cambridge, Cambridge CB2 1TN, United Kingdom }
\affiliation {IISER-Kolkata, Mohanpur, West Bengal 741252, India }
\affiliation {Rutherford Appleton Laboratory, HSIC, Chilton, Didcot, Oxon OX11 0QX, United Kingdom }
\affiliation {Whitman College, 345 Boyer Avenue, Walla Walla, WA 99362 USA }
\affiliation {National Institute for Mathematical Sciences, Daejeon 305-390, Korea }
\affiliation {Universit\'e de Lyon, F-69361 Lyon, France }
\affiliation {Hobart and William Smith Colleges, Geneva, NY 14456, USA }
\affiliation {Janusz Gil Institute of Astronomy, University of Zielona G\'ora, 65-265 Zielona G\'ora, Poland }
\affiliation {King's College London, University of London, London WC2R 2LS, United Kingdom }
\affiliation {Andrews University, Berrien Springs, MI 49104, USA }
\affiliation {Universit\`a di Siena, I-53100 Siena, Italy }
\affiliation {Trinity University, San Antonio, TX 78212, USA }
\affiliation {University of Washington, Seattle, WA 98195, USA }
\affiliation {Kenyon College, Gambier, OH 43022, USA }
\affiliation {Abilene Christian University, Abilene, TX 79699, USA }

  }{
    \author{The LIGO Scientific Collaboration}
    \affiliation{LSC}
    \author{The Virgo Collaboration}
    \affiliation{Virgo}
  }
}

\begin{abstract}
\iftoggle{endauthorlist}{
}{
  \newpage
}
We report results of a deep all-sky search for periodic gravitational waves from isolated neutron stars in data from the S6 LIGO science run. The search was possible thanks to the computing power provided by the volunteers of the \EatH distributed computing project. We find no significant signal candidate and set the most stringent upper limits to date on the amplitude of gravitational wave signals from the target population.
At the frequency of best strain sensitivity, between $170.5$ and $171$~Hz we set a 90\%\ confidence upper limit of $\sci{5.5}{-25}$, while at the high end of our frequency range, around 505 Hz, we achieve upper limits $\simeq {10}^{-24}$. At $230$ Hz we can exclude sources with ellipticities greater than $10^{-6}$ within 100 pc of Earth with fiducial value of the principal moment of inertia of $10^{38} \textrm{kg m}^2$. If we assume a higher (lower) gravitational wave spindown we constrain farther (closer) objects to higher (lower) ellipticities.\\
\end{abstract}

\maketitle

\author{The LIGO Scientific Collaboration and the Virgo Collaboration}
\thanks{Full author list given at the end of the article.}




%
%
\maketitle

\section{Introduction}
\label{sec:introduction}

In this paper we report the results of a deep all-sky \EatH \cite{EaHweb} search for continuous, nearly monochromatic gravitational waves (GWs) in data from LIGO's sixth science (S6) run. A number of all-sky searches  have been carried out on LIGO data, \cite{S6Powerflux, Aasi:2015rar, S5GC1HF, FullS5EH,FullS5Semicoherent, S5EH, EarlyS5Paper,S4IncoherentPaper,S4EH,S2FstatPaper}, of which \cite{S4EH, S5EH, FullS5EH} also ran on \EatH. The search presented here covers frequencies from 50~Hz through 510~Hz and frequency derivatives from {\paramfdothi}  through \paramfdotlo. In this range we establish the most constraining gravitational wave amplitude upper limits to date for the target signal population.

\section{LIGO interferometers and the data used}
\label{sec:S6intro} 

The LIGO gravitational wave network consists of two observatories, one in Hanford (WA) and the other in Livingston (LA) separated by a 3000-km baseline \cite{LIGO_detector}. The last science run (S6) \cite{LIGO:2012aa} of this network before the upgrade towards the advanced LIGO configuration \cite{TheLIGOScientific:2014jea} took place between July 2009 and October 2010. The analysis in this paper uses a subset of this data: from GPS 949469977  (2010 Feb 6  05:39:22 UTC) through GPS 971529850 (2010 Oct 19  13:23:55 UTC), selected for good strain sensitivity \cite{Shaltev}.  Since interferometers sporadically fall out of operation (``lose lock'') due to environmental or instrumental disturbances or for scheduled maintenance periods, the data set is not contiguous and each detector has a duty factor of about 50\% \cite{detchar2}. 
 
As done in \cite{S5EH}, frequency bands known to contain spectral disturbances have been removed from the analysis. Actually, the data has been substituted with Gaussian noise with the same average power as that in the neighbouring and undisturbed bands. 
Table \ref{A:cleanedBands} identifies these bands.

\section{The Search}

The search described in this paper targets nearly monochromatic gravitational wave signals as described for example by Eqs. 1-4 of \cite{S5EH}. Various emission mechanisms could generate such a signal as reviewed in  Section IIA of \cite{S2FstatPaper}. In interpreting our results we will consider a spinning compact object with a fixed, non-axisymmetric mass quadrupole, described by an ellipticity $\epsilon$.

We perform a stack-slide type of search using the GCT (Global correlation transform) method \cite{Pletsch:2008,Pletsch:2010}. In a stack-slide search the data is partitioned in segments and each segment is searched with a matched-filter method \cite{cutler}. The results from these coherent searches are combined by summing (stacking) the detection statistic values from the segments (sliding), one per segment ($\F_i$), and this determines the value of the core detection statistic: 
\begin{equation}
\label{eq:avF}
\avF:={1\over\Nseg} \sum_{i=1}^{\Nseg} \F_i.
\end{equation}
There are different ways to combine the single-segment $\F_i$ values, but independently of the way that this is done, this type of search is usually referred to as a ``semi-coherent search''. So stack-slide searches are a type of semi-coherent search. Important variables for this type of search are: the coherent time baseline of the segments $\Tcoh$, the number of segments used $\Nseg$, the total time spanned by the data $\Tobs$, the grids in parameter space and the detection statistic used to rank the parameter space cells. For a stack-slide search in Gaussian noise, $\Nseg\times 2\avF$ follows a $\chi^2_{4\Nseg}$ chi-squared distribution with $4\Nseg$ degrees of freedom.
\begin{figure}[h!tbp]
\begin{centering}
  \includegraphics[width=1.0\columnwidth]{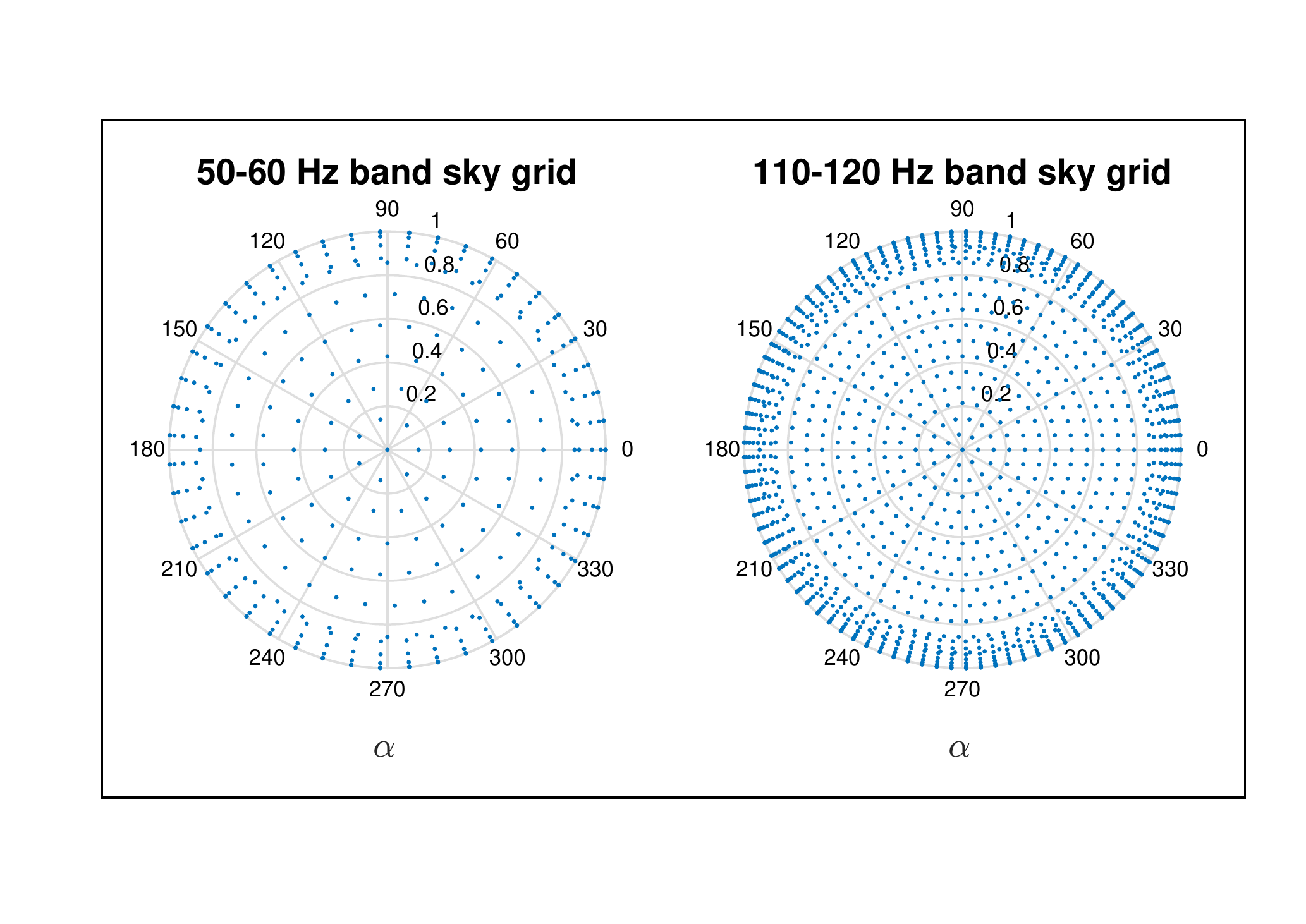}
\caption{Polar plots ($r,\theta$ plots with $\theta=\alpha$ and $r=\cos\delta$) of the grid points in the northern equatorial hemisphere sky for the band 50-60Hz (left panel) and for the band 110-120Hz (right panel). $\alpha$ is the right ascension coordinate and $\delta$ the declination coordinate. One can clearly see the higher density in the $-0.5\leq \delta \leq 0.5$ equatorial region and the higher density ($\propto {{f^2}}$) of grid points at higher frequencies. The southern hemispheres looks practically identical to the respective northern ones.}  
\label{fig:SkyGrids}
\end{centering}
\end{figure}
These parameters are summarised in Table \ref{tab:GridSpacings}. The grids in frequency and spindown are each described by a single parameter, the grid spacing, which is constant over the search range. The same frequency grid spacings are used for the coherent searches over the segments and for the incoherent summing. The spindown spacing for the incoherent summing, $\delta{\dot{f}}$, is finer than that used for the coherent searches, $\delta{\dot{f_c}}$, by a factor $\gamma$. The notation used here is consistent with that used in previous observational papers \cite{GalacticCenterSearch} and in the GCT methods papers cited above. 

\begin{figure}[h!tbp]
   \includegraphics[width=\columnwidth]{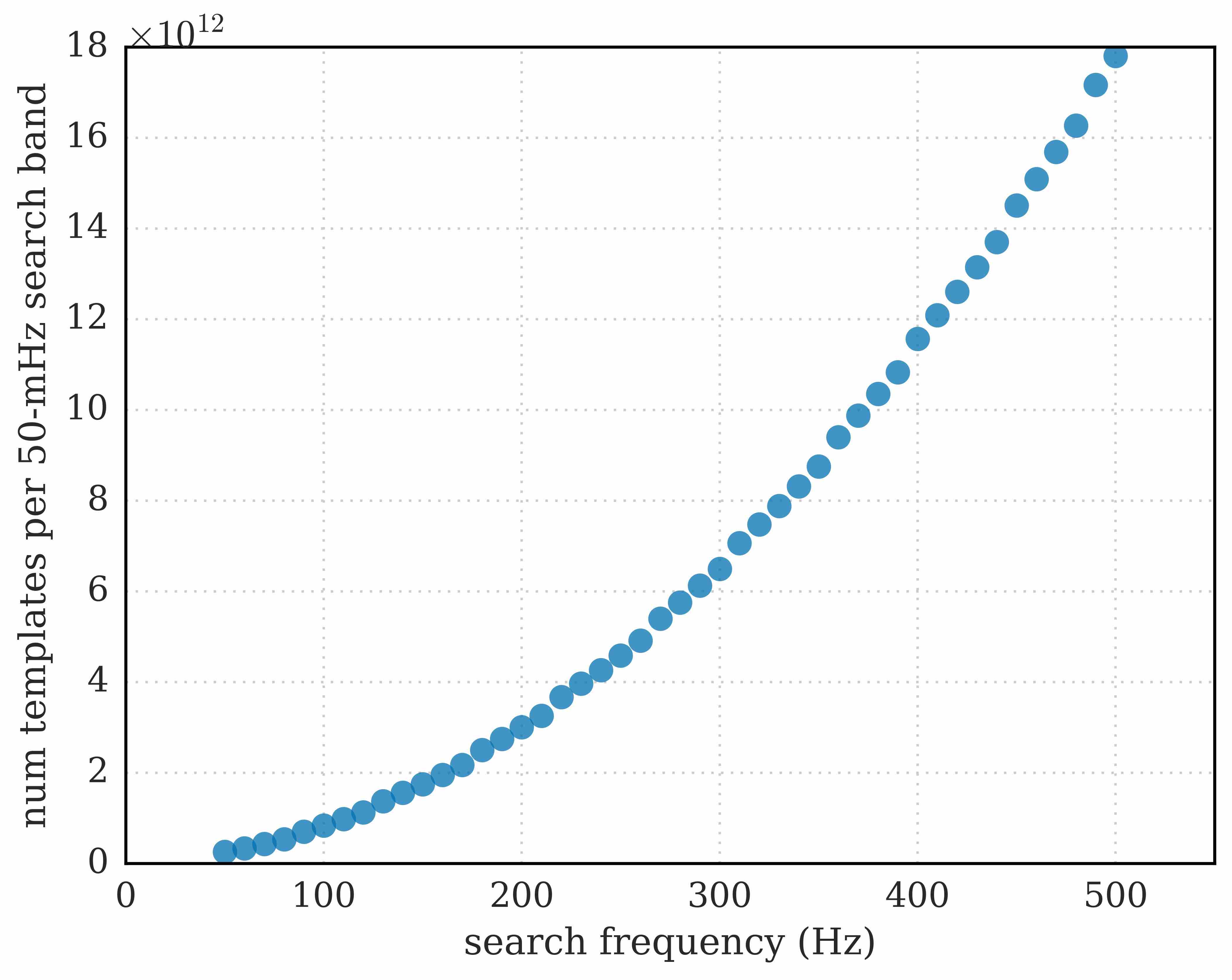}
\caption{Number of searched templates in 50-mHz bands. The variation with frequency is due to the increasing sky resolution.  
$N_f \times N_{\dot{f}} \sim 3.7\times 10^{8}$, where $N_f$ and $N_{\dot{f}}$ are the number of $f$ and $\dot{f}$ templates searched in 50-mHz bands. The total number of templates searched between 50 and 510 Hz is 
{\paramtotaltemplates}. 
}  
\label{fig:NumberOfTemplatesIn50mHz}
\end{figure}

The sky grid is the union of two grids: one is uniform over the projection of the celestial sphere onto the equatorial plane, and the tiling (in the equatorial plane) is approximately square with sides of length
\begin{equation}
d(m_{\text{sky}})={1\over f}
{
{\sqrt{ m_{\text{sky}} } }
\over {\pi \tau_{E}} 
},
\label{eq:skyGridSpacing}
\end{equation} 
with $m_{\text{sky}}=0.3$ and $\tau_{E}\simeq0.021$ s being half of the light travel time across the Earth. As was done in \cite{S5EH}, the sky-grids are constant over 10 Hz bands and the spacings are the ones associated through Eq. \ref{eq:skyGridSpacing} to the highest frequency $f$ in the range. 
The other grid is limited to the equatorial region ($0\leq \alpha\leq 2\pi$ and $-0.5\leq \delta \leq 0.5$), with constant right ascension $\alpha$ and declination $\delta$ spacings equal to $d(0.3)$ -- see Fig.\ref{fig:SkyGrids}. The reason for the equatorial ``patching'' with a denser sky grid is to improve the sensitivity of the search: the sky resolution actually depends on the ecliptic latitude and the uniform equatorial grid under-resolves particularly in the equatorial region. The resulting number of templates used to search 50-mHz bands as a function of frequency is shown in Fig. \ref{fig:NumberOfTemplatesIn50mHz}. 


\begin{table*}[t]
\begin{tabular}{|c|c|}
\hline
\hline
 $\Tcoh$ & 60 hrs\\
 \hline
 $\Tref$  & 960499913.5 GPS sec \\
  \hline
$\Nseg$ & 90 \\
  \hline
$\delta f$ & $1.6\times 10^{-6}$ Hz \\
 \hline
 $\delta {\dot{f_c}}$ & $5.8 \times 10^{-11}$ Hz/s \\
  \hline
$\gamma$ & 230 \\
  \hline
$m_{\text{sky}}$ &0.3 + equatorial patch\\
 \hline
\hline
\end{tabular}

\caption{Search parameters rounded to the first decimal figure. $\Tref$ is the reference time that defines the frequency and frequency derivative values.
}
\label{tab:GridSpacings}
\end{table*}

The search is split into work-units (WUs) sized to keep the average \EatH volunteer computer busy for about {\paramWUcputimeHours} hours. Each WU searches a $50$ mHz band, the entire spindown range and {\paramWUavgskypointsInt} points in the sky, corresponding to {\paramWUtotaltemplates}~templates out of which it returns only the top 3000. A total of {\paramtotalWUsmillions}~million WUs are necessary to cover the entire parameter space. 
The total number of templates searched is \paramtotaltemplates.

\subsubsection{The ranking statistic}
The search was actually carried out in separate \EatH runs that used different ranking statistics to define the top-candidate-list, reflecting different stages in the development of a detection statistic robust with respect to spectral lines in the data \cite{Keitel:2013}. In particular, three ranking statistics were used: the average $2\F$ statistic over the segments, $2\avF$,  which in essence at every template point is the likelihood of having a signal with the shape given by the template versus having Gaussian noise; the line-veto statistic $\OSLsc$ which is the odds ratio of having a signal versus having a spectral line; and a general line-robust statistic, $\OSNsc$, that tests the signal hypothesis against a Gaussian noise + spectral line noise model. Such a statistic can match the performance of both the standard average $2\avF$ statistic in Gaussian noise and the line-veto statistic in presence of single-detector spectral disturbances and statistically outperforms them when the noise is a mixture of both \cite{Keitel:2013}.

We combine the $2\avF$-ranked results with the $\OSLsc$-ranked results to produce a single list of candidates ranked according to the general line-robust statistic $\OSNsc$. We now explain how this is achieved. Alongside the detection statistic value and the parameter space cell coordinates of each candidate, the \EatH application also returns the single-detector $2\avFX$ values (``$X$'' indicates the detector). These are used to compute, for every candidate of any run, the $\OSNsc$ through Eq. 61 of \cite{Keitel:2013}:
\begin{equation}
  \begin{split}
  \label{eq:logOSNsc}
  \ln &\OSNsc = \ln \oSLsc + \scF - \scFmaxG\\
  &- \ln \left( \eto{\scFth-\scFmaxG} + \avgX{\rsc^X \eto{\scFX-\scFmaxG} } \right)\,,
  \end{split}
\end{equation}
with the angle-brackets indicating the average with respect to detectors ($X$) and 
\begin{align}
  \label{eq:DefsforLogOSGL1}
&\scF=\Nseg \avF\\
&\scFX=\Nseg \avFX\\
&\scFmaxG \equiv \text{max}\left( \scFth,\,\scFX + \ln \scrX \right)\\
&\scFth \equiv \scFtho - \ln {\oLGsc}\\
\label{eq:cstar}
&\scFtho \equiv \ln \cF^{\Nseg} ~~{\text{with}}~\cF~{\text{set to }} 20.64\\
&\oLGsc  = \sum_X \oLGsc^X \\
&\rsc^X \equiv \frac{\oLGsc^X}{\oLGsc/\Ndet}\\
&\lineprobsc \equiv \frac{\oLGsc}{1 + \oLGsc}
\end{align}
where $\oLGsc^X$ is the assumed prior probability of a spectral line
occuring in any frequency bin of detector X, $\lineprobsc$ is the line prior
estimated from the data, $\Ndet = 2$ is the number of detectors, and $ \oSLsc$ is
an assumed prior probability of a line being a signal (set arbitrarily
to 1; its specific value does not affect the ranking statistic). Following the reasoning of Eq. 67 of \cite{Keitel:2013}, with $\Nseg=90$ we set $\cF=20.64$ corresponding to a Gaussian false-alarm probability  of $10^{-9}$ and an average $2\avF$ transition scale of $\sim 6$ ($\Ftho\sim 3$). The 
$ \oLGsc^X$ values are estimated from the data as described in Section VI.A of \cite{Keitel:2013} in 50-mHz bands with a normalized-SFT-power threshold $\Psftthr^X=\Psftthr(\pFA=10^{-9},\Nsft^X\sim 6000)\approx 1.08$. For every 50-mHz band the list of candidates from the $2\avF$-ranked run is merged with the list from the $\OSLsc$-ranked run and duplicate candidates are considered only once. The resulting list is ranked by the newly-computed $\OSNsc$ and the top 3000 candidates are kept. This is our result-set and it is treated in a manner that is very similar to \cite{S5GC1HF}.


\subsubsection{Identification of undisturbed bands}
\label{sec:visualInspection}
Even after the removal of disturbed data caused by spectral artefacts of known origin, the statistical properties of the results are not uniform across the search band. In what follows we concentrate on the subset of the signal-frequency bands having reasonably uniform statistical properties. This still leaves us with the majority of the search parameter space while allowing us to use methods that rely on theoretical modelling of the significance in the statistical analysis of the results.  
Our classification  of ``clean" vs. ``disturbed" bands has no pretence of being strictly rigorous, because strict rigour here is neither useful nor practical. The classification serves the practical purpose of discarding from the analysis regions in parameter space with evident disturbances and must not dismiss real signals. The classification is carried out in two steps: a visual inspection and a refinement on the visual inspection. 

\begin{figure}[h!tbp]
  \includegraphics[width=\columnwidth]{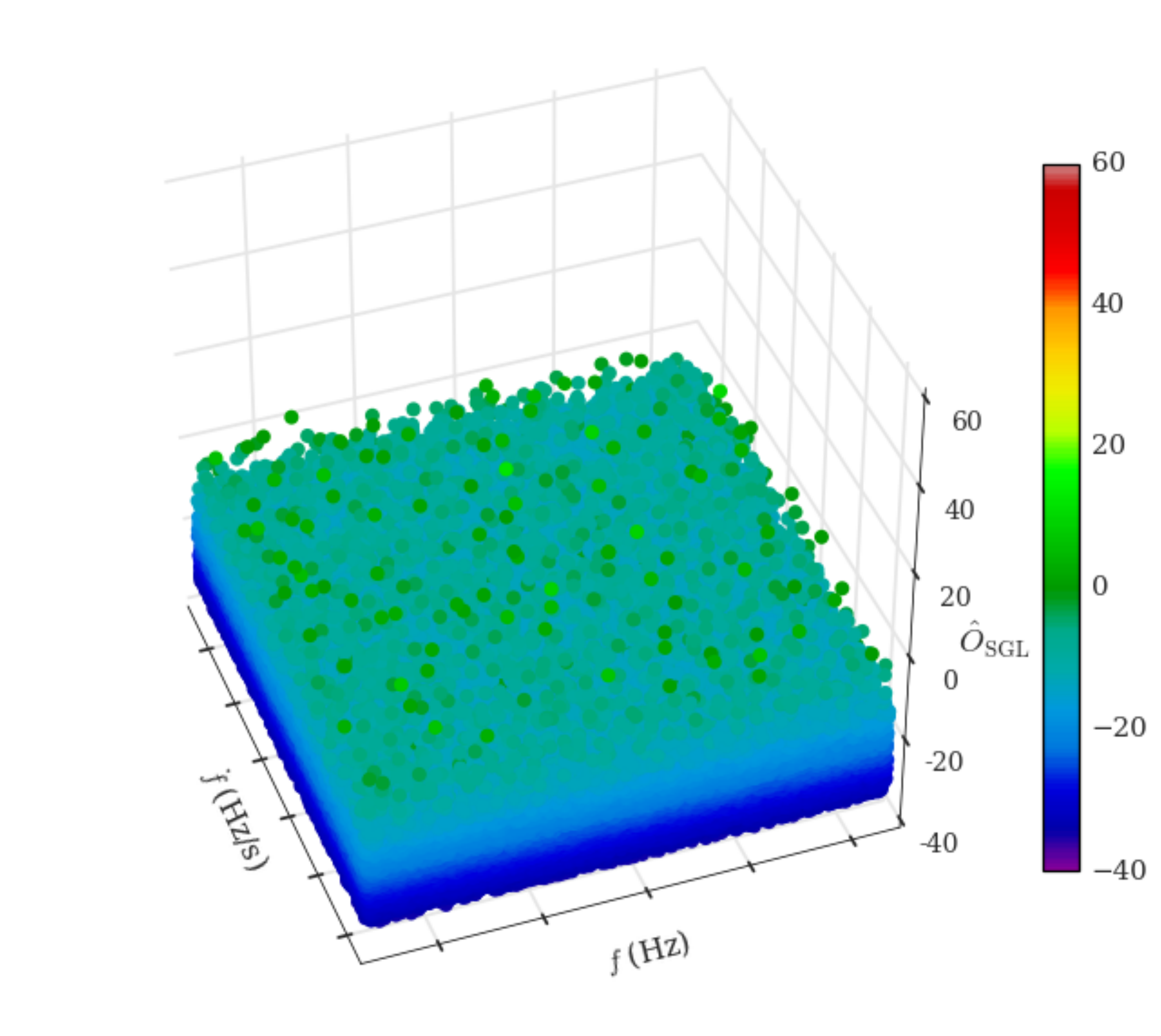}
  \includegraphics[width=\columnwidth]{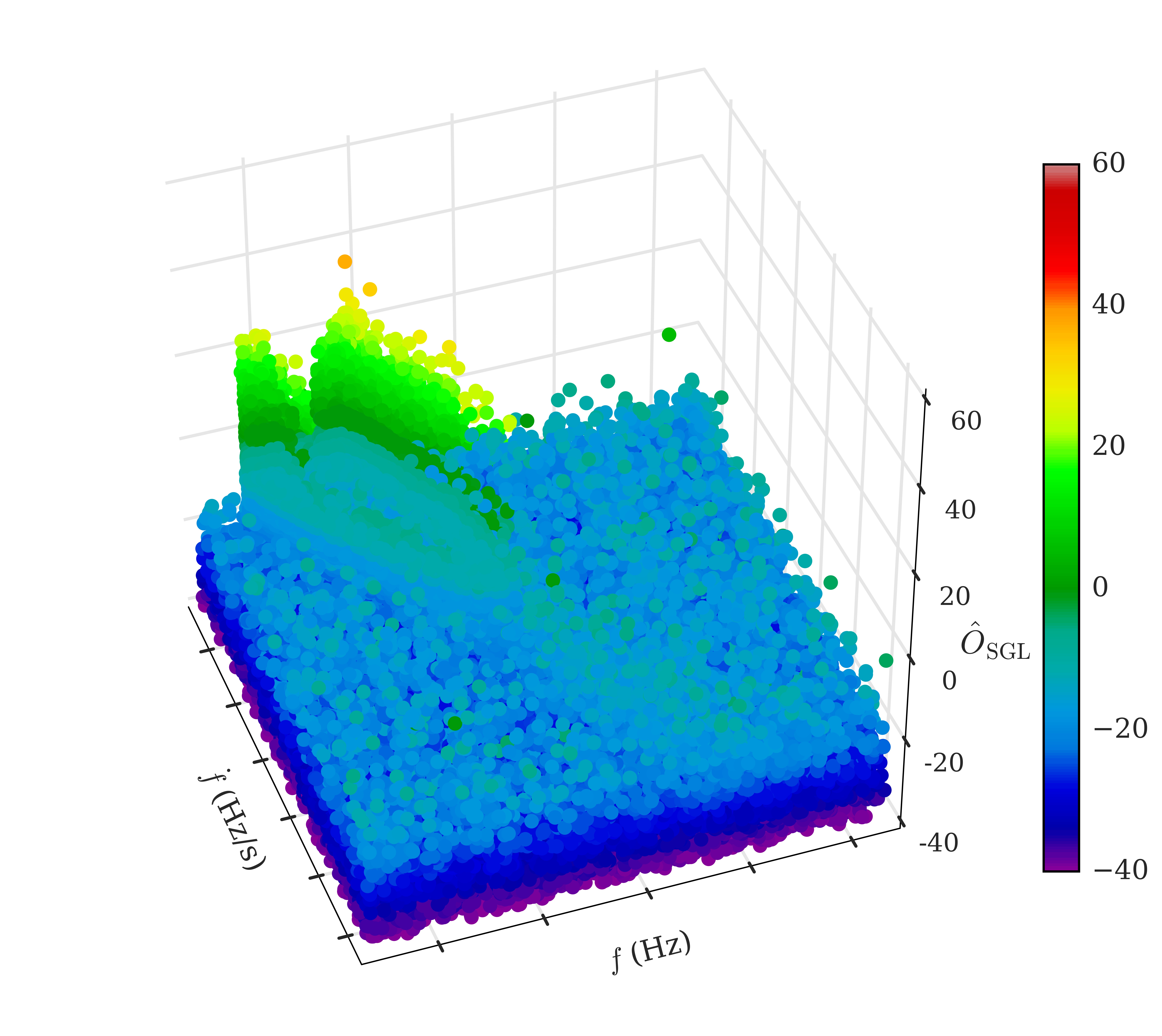}
   \includegraphics[width=\columnwidth]{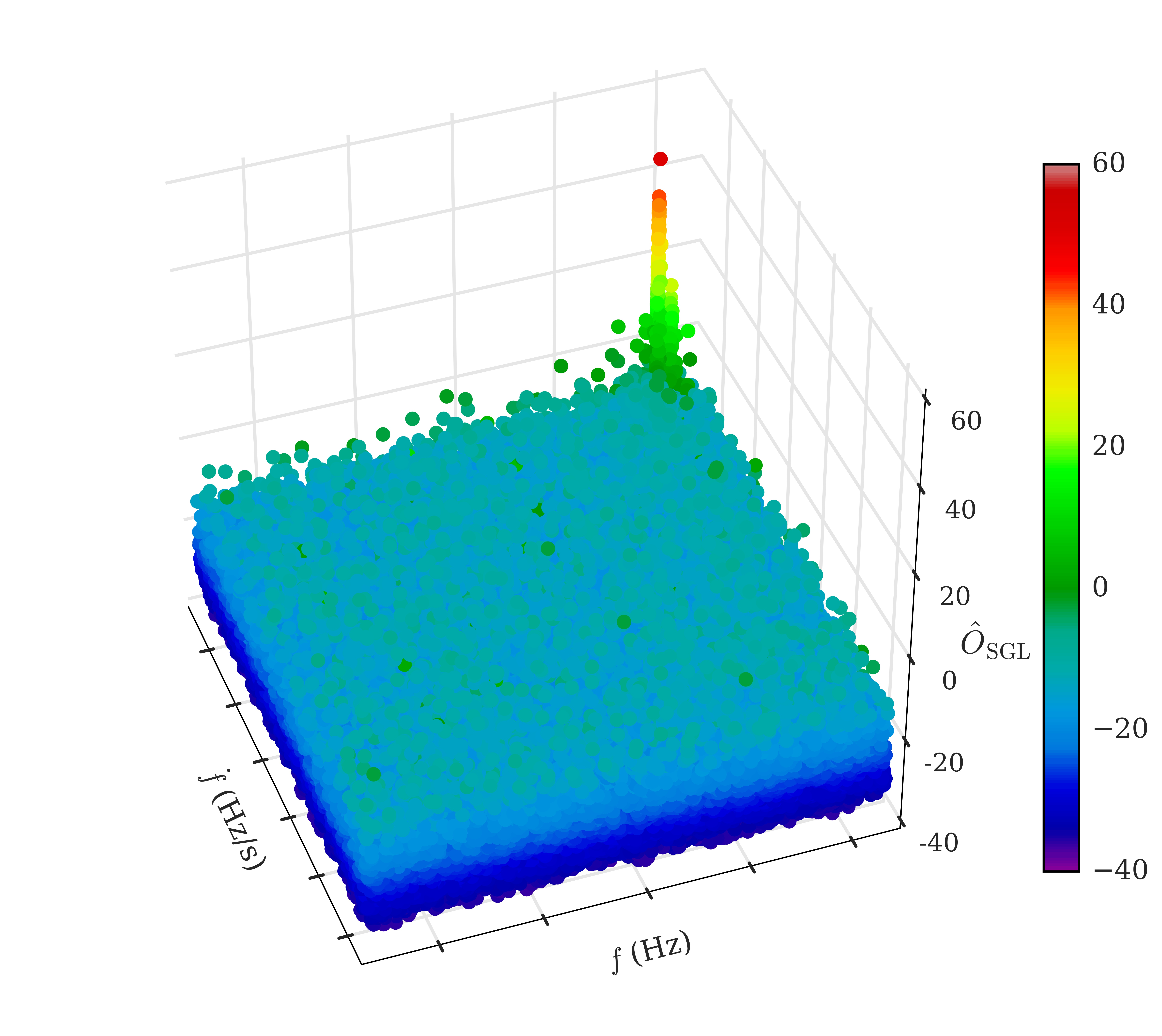}
\caption{On the z-axis and color-coded is the $\OSNsc$ in three 50-mHz bands. The top band was marked as ``undisturbed". The middle band is an example of a ``disturbed band". The bottom band is an example of an ``undisturbed band" but containing a signal, a fake one, in this case.} 
\label{fig:VIGquietdisturbed}
\end{figure}

The visual inspection is performed by three scientists who each look at various distributions of the detection statistics over the entire sky and spindown parameter space in 50-mHz bands. They rank each band with an integer score 0,1,2,3 ranging from ``undisturbed" (0) to ``disturbed" (3) . A band is considered ``undisturbed" if all three rankings are 0. The criteria agreed upon for ranking are that the distribution of detection statistic values should not show a visible trend affecting a large portion of the $f-\dot{f}$ plane and, if outliers exist in a small region, outside this region the detection statistic values should be within the expected ranges. Fig. \ref{fig:VIGquietdisturbed} shows the $\OSNsc$ for three bands: two were marked as undisturbed and the other as disturbed. One of the bands contains the $f-\dot{f}$ parameter space that harbours a fake signal injected in the data to verify the detection pipelines. The detection statistic is elevated in a small region around the signal parameters. The visual inspection procedure does not mark as disturbed bands with such  features.

Based on this visual inspection 13\% 
of the bands between 50 and 510 Hz are marked as ``disturbed". Of these, 34\% 
were given by all visual inspectors rankings smaller than 3, i.e. they were only marginally disturbed. Further inspection  ``rehabilitated'' 42\% of these. 
 As a result of this refinement in the selection procedure we exclude from the current analysis 11\% of the searched frequencies.

 \begin{figure}[h!tbp] 
\includegraphics[width=\columnwidth]{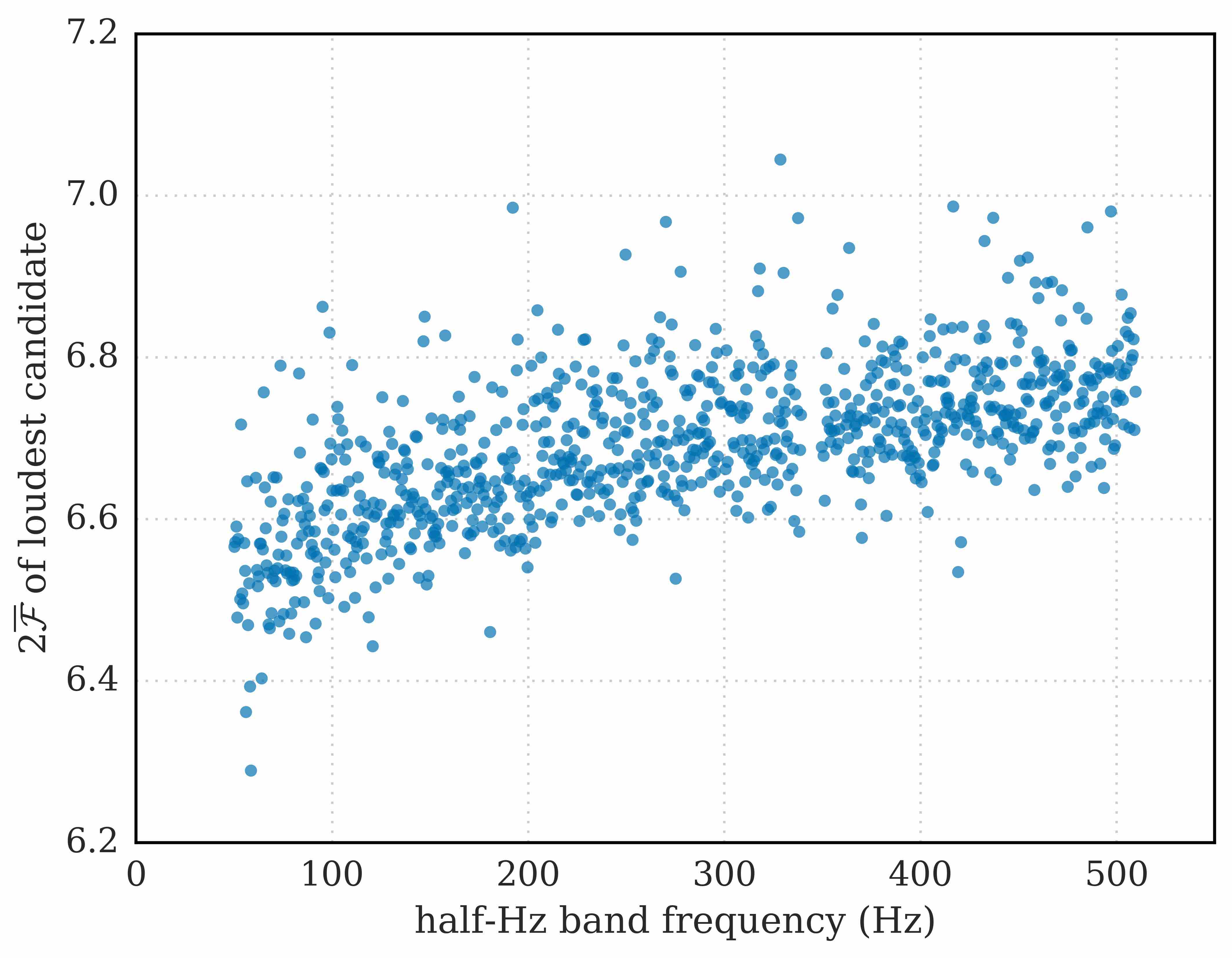}

        \caption{Highest values of $2\avF$ in every half-Hz band as a function of band frequency. Since the number of templates increases with frequency so does the loudest $2\avF$.}  
\label{fig:loudestHalfHzVersusFreq}
\end{figure}
\begin{figure}[h!tbp]
    \includegraphics[width=\columnwidth]{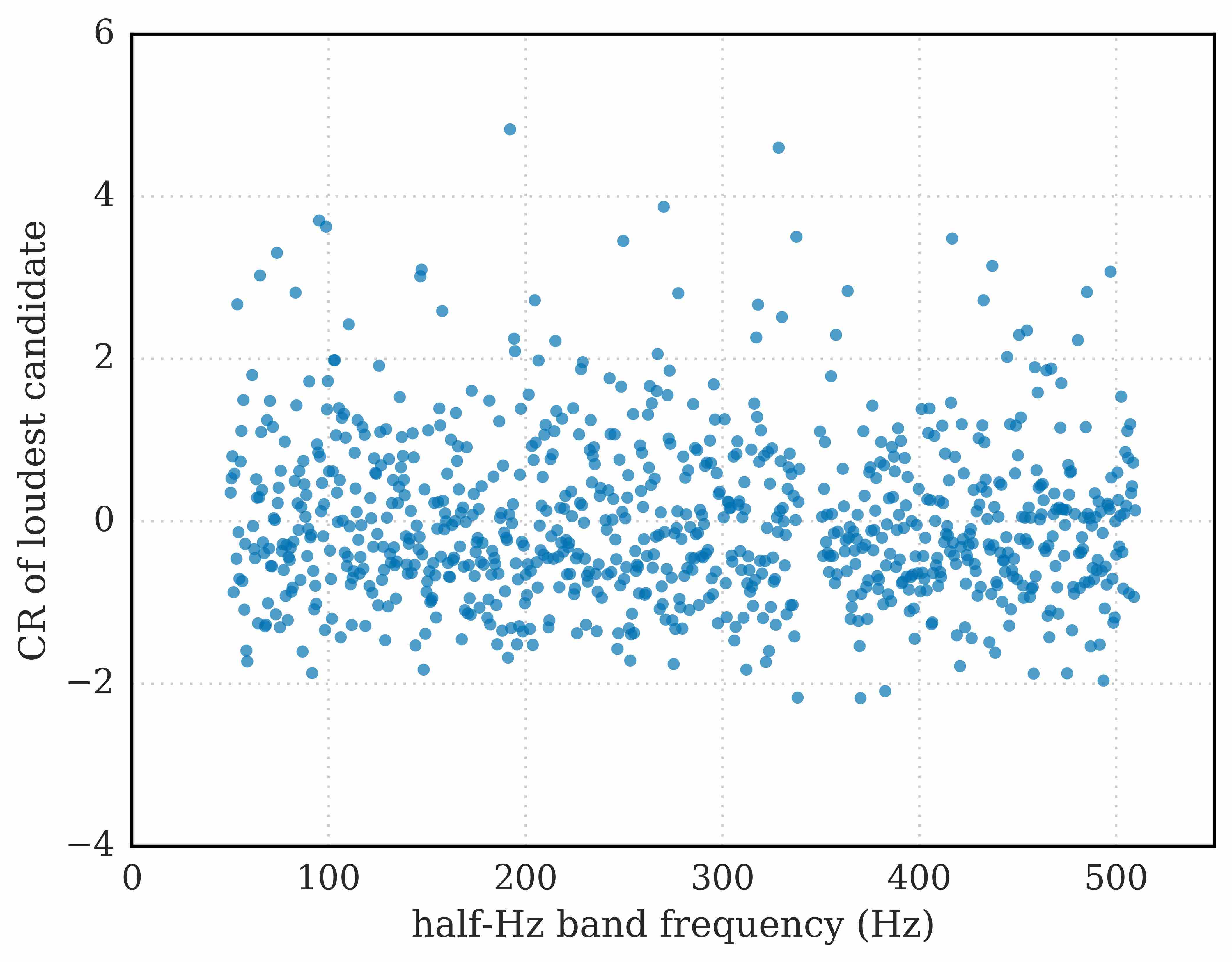}
     \caption{Highest values of the significance (CR) in every half-Hz band as a function of band frequency. Since the significance folds in the expected value for the loudest  $2\avF$ and its standard deviation, the significance of the loudest in noise does not increase with frequency. Our results are consistent with this expectation.}  
\label{fig:loudestSigmaHalfHzVersusFreq}
\end{figure}
\begin{figure}[h!]
     \includegraphics[width=\columnwidth]{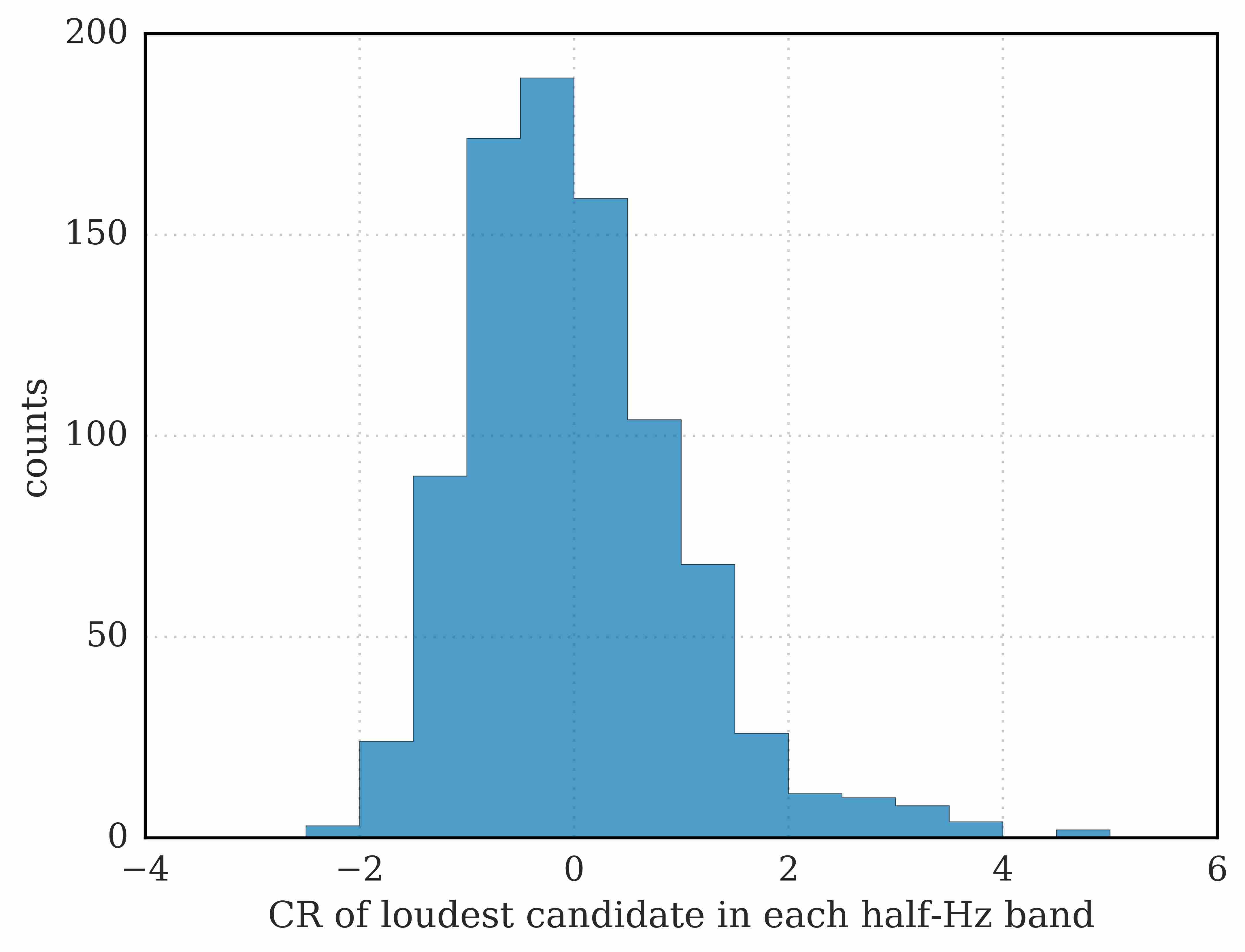}
     \caption{Histogram of the highest values of the significance CR in every half-Hz band.}  
\label{fig:loudestSigmaHistHalfHzVersusFreq}
\end{figure}

\begin{figure}[h!]
     \includegraphics[width=\columnwidth]{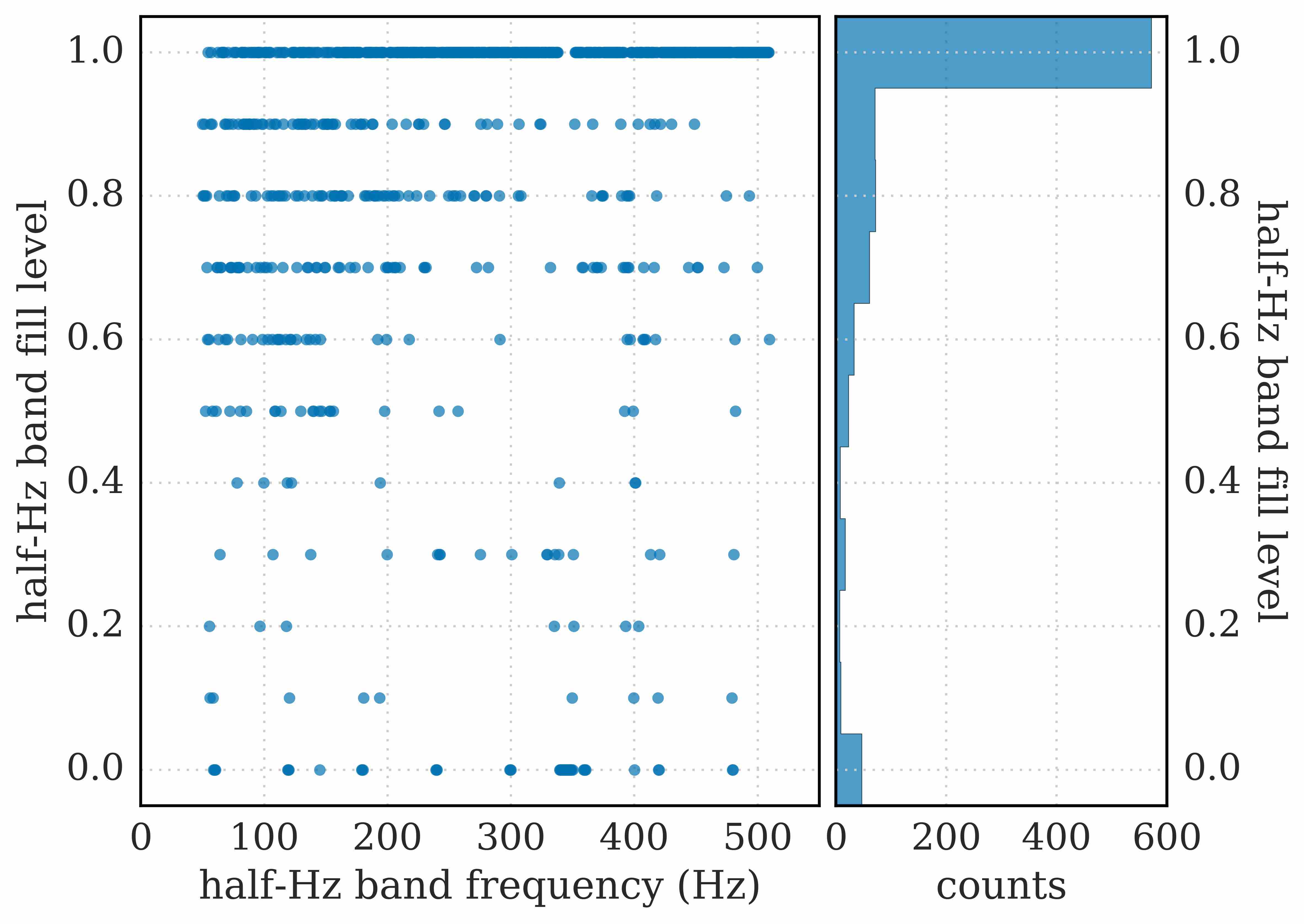}
     \caption{The fraction of 50mHz bands (in signal frequency) which contribute to the results in every half-Hz band. As explained in the text, some bands are excluded because they are all from fake data or because they are marked as disturbed by the visual inspection. The list of excluded bands is given in Table \ref{A:excluded50mHzBands}.}  
\label{fig:fill-level}
\end{figure}

Fig. \ref{fig:loudestHalfHzVersusFreq} shows the highest values of the detection statistic in half-Hz signal-frequency bands compared to the expectations. The set of candidates that the highest detection statistic values are picked from,  does not include the 50mHz signal-frequency bands that stem entirely from fake data, from the cleaning procedure, or that were marked as disturbed. In this paper we refer to the candidates with the highest value of the detection statistic as the {\it{loudest}} candidates. 

The loudest expected value over $N_{\text{trials}}$ independent trials of $2\avF$ 
is determined\footnote{After a simple change of variable from $2\avF$  to $\Nseg\times 2\avF$.} by numerical integration of the probability density function given, for example, by Eq. 7 of \cite{GalacticCenterSearch}. For this search, we estimate that $N_{\text{trials}}\simeq 0.87 \,N_{\text{templ}}$, with $N_{\text{templ}}$ being the number of templates searched.

As a uniform measure of significance of the highest $2\avF$ value across bands that were searched with different values of $N_{\text{trials}}$ we introduce the critical ratio CR defined as the deviation of the measured highest $2\avF$ from the expected value, measured in units of the standard deviation:
\begin{equation}
\label{eq:CR}
{\text{CR}}:={{2\avF_{\text{meas}}- 2\avF_{\text{exped}}}\over{\sigma_{\text{exped}}}}.
\end{equation}

The highest and most significant detection statistic value from our search is $2\avF=8.6$ at a frequency of about 52.76 Hz with a CR=29. This is due to a fake signal. The second highest value of the detection statistic is 7.04 at a frequency of about 329.01 Hz corresponding to a CR of 4.6. The second highest-CR candidate has a $2\avF$ of 6.99, is at 192.16 Hz and has a CR=4.8. 

Sorting loudest candidates from half-Hz bands according to detection statistic values is not the same as sorting them according to CR. The reason for this is that the number of templates is not the same for all half-Hz bands. This is due to the grid spacings decreasing with frequency (Eq. \ref{eq:skyGridSpacing}) and to the fact that, as previously explained, some 50-mHz bands have been excluded from the current analysis and hence some half-Hz bands comprise results from fewer than ten 50mHz bands. Fig.\ref{fig:fill-level} gives the fill-level of each half-Hz band, i.e. how many 50mHz bands have contributed candidates to the analysis out of ten. We use the CR as a measure of the significance because it folds in correctly the effect of varying number of templates in the half-Hz bands. 

\begin{figure}[h!]
     \includegraphics[width=\columnwidth]{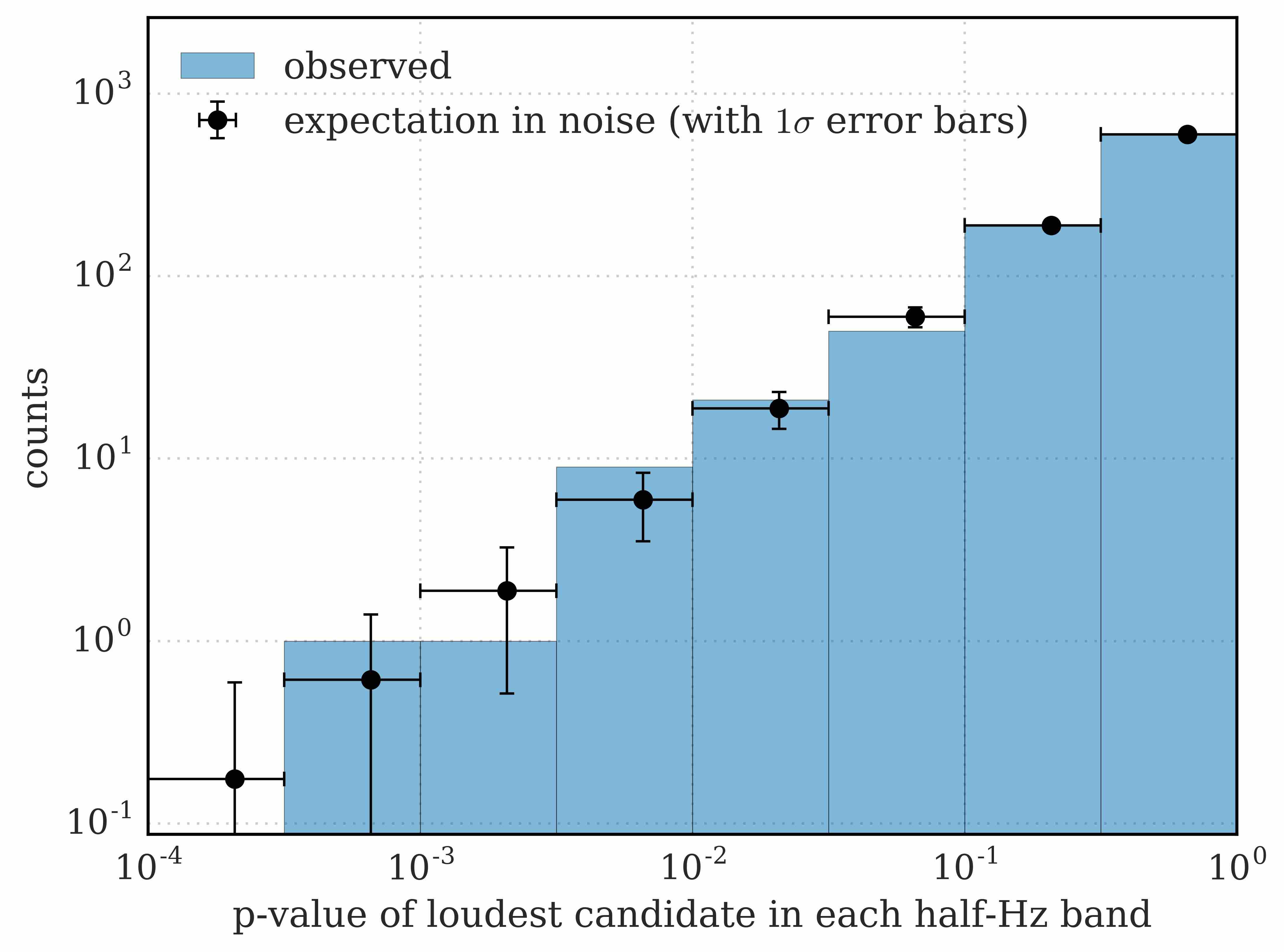}
     \caption{p-values for the loudest in half-Hz bands of our data (histogram bars) and expected distribution of pure noise data for reference (black markers).}  
\label{fig:pvalues}
\end{figure}

After excluding the candidate due to the fake signal, in this data we see no evidence of a signal: the distribution of p-values associated with every measured half-Hz band loudest is consistent with what we expect from noise-only across the measured range (Fig.\ref{fig:pvalues}). In particular we note two things: 1) the two candidates at CR=4.6 and CR=4.8 are not significant when we consider how many half-Hz bands we have searched and 2) there is no population of low significance candidates deviating from the expectation of the noise-only case. The p-value for the loudest measured in any half-Hz band searched with an effective number of independent trials $N_{\text{trials}}=0.87 ~N_{\text{trials}}$ is obtained by integrating Eq. 6 of \cite{GalacticCenterSearch} between the observed value and infinity.

\section{Upper limits}
\label{sec:upper limits}

\begin{figure*}
 \includegraphics[width=0.8\textwidth]{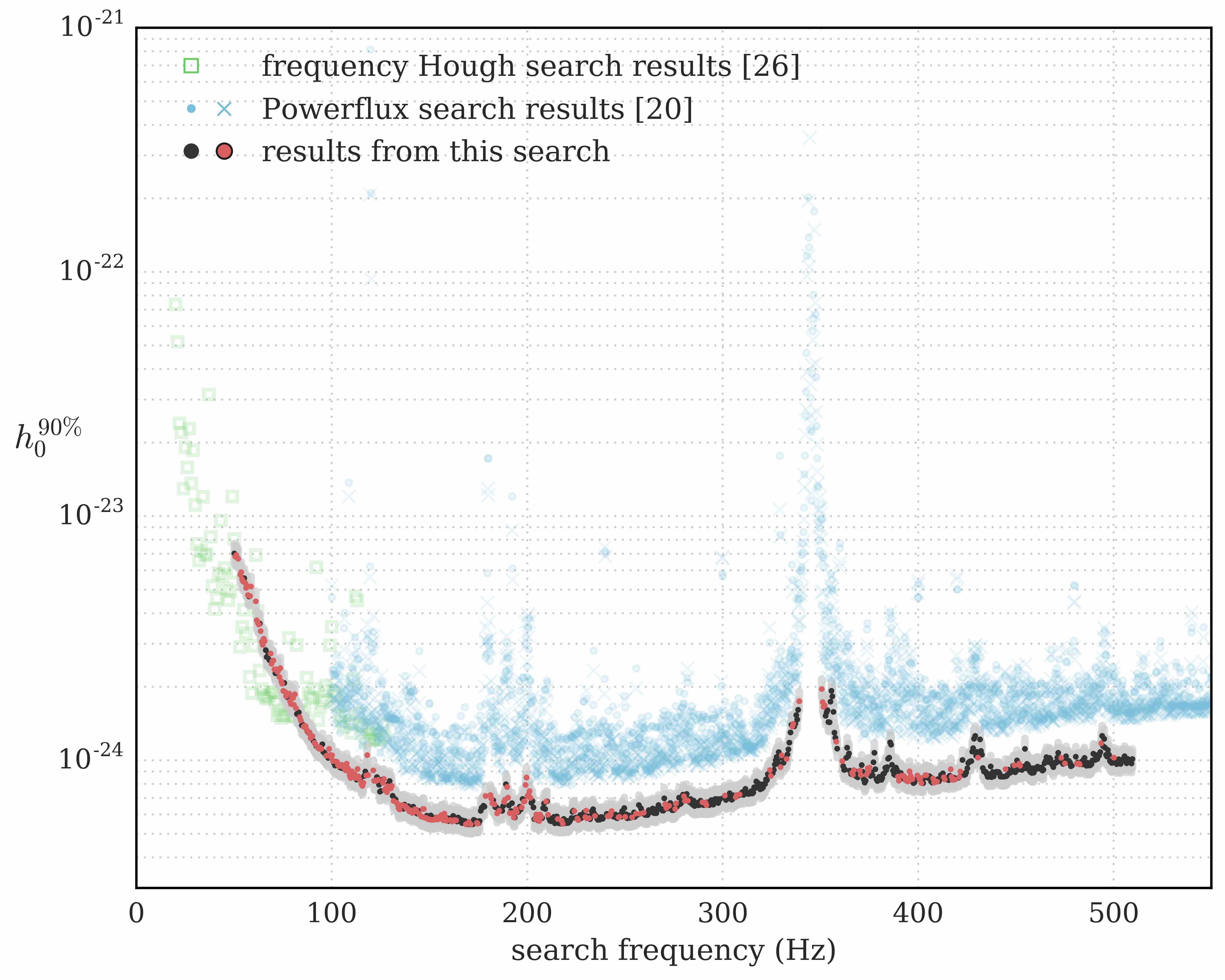}
\caption{90\% confidence upper limits on the gravitational wave amplitude of signals with frequency within half-Hz bands, from the entire sky and within the spindown range of the search. The light red markers denote half-Hz bands where the upper limit value does not hold for all frequencies in that interval. A list of the excluded frequencies is given in the Appendix. Although not obvious from the figure, due to the quality of the data we were not able to analyse the data in some half-Hz bands, so there are some points missing in the plot. For reference we also plot the upper limit results from two searches: one on the same data (Powerflux) \cite{S6Powerflux} and on contemporary data from the Virgo detector (frequency Hough) \cite{Aasi:2015rar}. The Powerflux points are obtained by rescaling the best (crosses) and worst-case (dots) upper limit values as explained in the text. It should be noted that the Powerflux upper limits are set at $95\%$ rather than $90\%$ but refer to 0.25 Hz bands rather than half-Hz.}  
\label{fig:ULs}
\end{figure*}
\begin{figure*}
\includegraphics[width=0.8\textwidth]{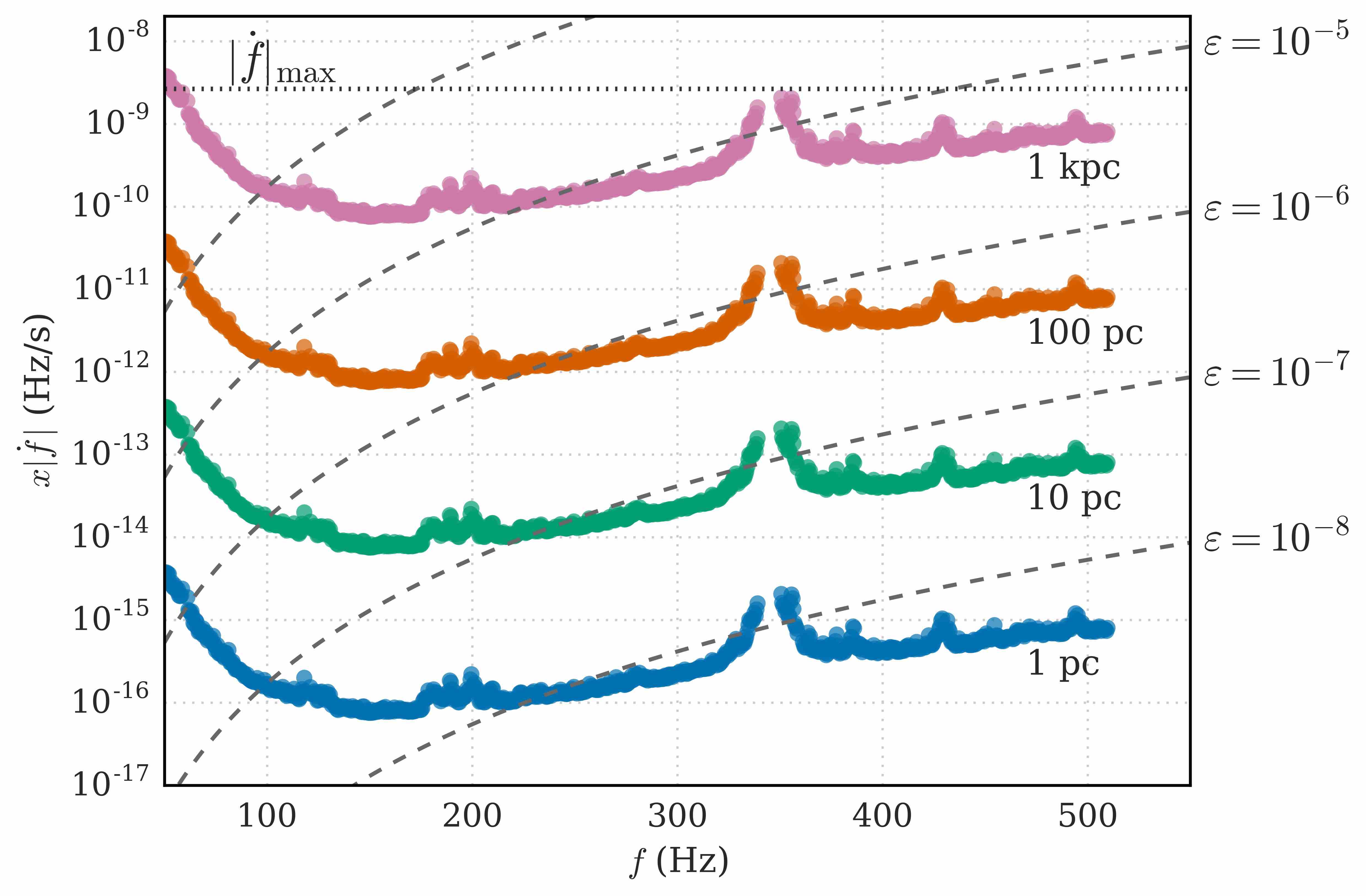}
\caption{Gravitational wave amplitude upper limits recast as curves in the $f-x|{\dot{f}}|$ plane for sources at given distances and having assumed $I=10^{38}~{\textrm{kg m}^2}$. $f$ is the signal frequency and $x|{\dot{f}}|$ is the gravitational-wave spindown, i.e. the fraction of the actual spindown that accounts for the rotational energy loss due to GW emission. Superimposed are curves of constant ellipticity $\epsilon(f,\dot{f} | I=10^{38}~{\textrm{kg m}^2}$). The dotted line at $|{\dot{f}}|_\textrm{max}$ indicates the maximum magnitude of searched spindown.}  
\label{fig:reach}
\end{figure*}

The search did not reveal any continuous gravitational wave signal in the parameter volume that was searched. We hence set frequentist upper limits on the maximum gravitational wave amplitude consistent with this null result in  half-Hz bands : $h_0^{90\%}$(f). $h_0^{90\%}$(f) is the GW amplitude such that 90\% of a population of signals with parameter values in our search range would have produced a candidate louder than what was
observed by our search. This is the criterion hereafter referred to as ``detection".

Evaluating these upper limits with injection-and-recovery Monte Carlo simulations in every half-Hz band is too computationally intensive. So we perform them in a subset of 50 bands and infer the upper limit values in the other bands from these. The 50 bands are evenly spaced in the search frequency range. 
For each band $j=1\dots 50$, we  
measure the 90\% upper limit value corresponding to different detection criteria. The different detection criteria are defined by different CR values for the assumed measured loudest. The first CR bin, $\CR_0$, is for CR values equal to or smaller than $0$, the next bins are for $i < \CR_i \leq (i+1)$  with $i=1 \ldots 5$. Correspondingly we have  ${h_{0,\CR_i}^{90\%,j}}$ for each band. For every detection criteria and every band we determine the sensitivity depth \cite{GalacticCenterMethod}, and by averaging these sensitivity depths over the bands we derive a sensitivity depth for every detection criteria: ${\mathcal{D}}_{\CR_i}^{90\%}={1/50}\sum_j {\mathcal{D}}_{\CR_i}^{90\%,j}$. We use these to set upper limits in the bands $k$ where we have not performed injection-and-recovery simulations as
\begin{equation}
{ h{_{0}^{90\%}}(f_k) } = 
{
{\sqrt{S_h(f_k)}} \over { 
{{\mathcal{D}}}_{\CR_i(k)}^{90\%} 
}
},
\label{eq:ULfromSensDepth}
\end{equation}
where $\CR_i(k)$ is the significance bin of the loudest candidate of the $k^{\text{th}}$ band and $S_h(f_k)$ the power spectral density of the data (measured in $1/\sqrt{\text{Hz}}$). The values of the sensitivity depths range between ${{\mathcal{D}}}_{\CR_6}^{90\%} \simeq 33 ~(1/\sqrt{\text{Hz}})$ and ${{\mathcal{D}}}_{\CR_0}^{90\%}\simeq 37 ~(1/\sqrt{\text{Hz}})$. The uncertainties on the upper limit values introduced by this procedure are $\simeq 10\%$ of the nominal upper limit value. We represent this uncertainty as a shaded region around the upper limit values in Fig.\ref{fig:ULs}. The upper limit values are also provided in tabular form in the Appendix in Table IV. We do not set upper limits in half-Hz bands where the results are entirely  produced with fake data inserted by the cleaning procedure described in Section \ref{sec:S6intro}. 
Upper limits for such bands will not appear in Table \ref{A:ULs} nor in Fig. \ref{fig:ULs}. There also exist 50-mHz bands that include contributions from fake data as a result of the cleaning procedure or that have been excluded from the analysis because they were marked as disturbed by the visual inspection procedure described in Section \ref{sec:visualInspection}. We mark the half-Hz bands which host these 50mHz bands with a different colour (light red) in Fig.\ref{fig:ULs}. In Table \ref{A:excluded50mHzBands} in the Appendix we provide a complete list of such 50-mHz bands because the upper limit values do not apply to those 50-mHz bands. Finally we note that, due to the cleaning procedure, there exist signal frequency bands where the search results {\it{might}} have contributions from fake data. We list these signal-frequency ranges in Table \ref{A:suspectFreqRanges}. For completeness this table also contains the cleaned bands of Table \ref{A:excluded50mHzBands}, under the column header ``all fake data''.

\section{Conclusions}

Our upper limits are the tightest ever placed for this set of target signals. The smallest value of the GW amplitude upper limit is $5.5\times 10^{-25}$ in the band 170.5-171 Hz. Fig. \ref{fig:ULs} shows the upper limit values as a function of search frequency. We also show the upper limits from \cite{S6Powerflux}, another all-sky search on S6 data, rescaled according to \cite{WetteSensEst:2012} to enable a direct comparison with ours. Under the assumption that the sources are uniformly distributed in space, our search probes a volume in space a few times larger than that of \cite{S6Powerflux}.  It should however be noted that 
\cite{S6Powerflux} examines a much broader parameter space than the one presented here.
The Virgo VSR2 and VSR4 science runs were contemporary to the S6 run and more sensitive at low frequency with respect to LIGO. The Virgo data were analysed in search of continuous signals from the whole sky in the frequency range 20\,Hz - 128\,Hz and a narrower spindown range than that covered here, with $|{\dot{f}}| \leqslant 10^{-10}$ Hz/s \cite{Aasi:2015rar}. Our sensitivity is comparable to that achieved by that search and improves on it above 80 Hz.

Following \cite{Ming:2015jla}, we define the fraction $x$ of the spindown rotational energy emitted in gravitational waves. The star's ellipticity necessary to sustain such emission is
\begin{equation}
\label{eq:GWspindown}
\epsilon(f,x{\dot{f}})=\sqrt{{5c^5\over 32\pi^4 G}{x{\dot{f}}\over If^5}},
\end{equation}
where $c$ is the speed of light, $G$ is the gravitational constant, $f$ is the GW frequency and $I$ the principal moment of inertia of the star. Correspondingly, $x{\dot{f}}$ is the spindown rate that accounts for the emission of GWs and this is why we refer to it as the GW spindown. The gravitational wave amplitude $h_0$ at the detector coming from a GW source like that of Eq.\ref{eq:GWspindown}, at a distance $D$ from Earth is
\begin{equation}
\label{eq:h0}
h_0(f,x\dot{f},D)={1\over D}\sqrt{{5GI\over 2c^3}{x{\dot{f}}\over f}}.
\end{equation}
Based on this last equation, we can use our GW amplitude upper limits to bound the minimum distance for compact objects emitting continuous gravitational waves under different assumptions on the object's ellipticity (i.e. gravitational wave spindown). This is shown in Fig. \ref{fig:reach}. We find that for most frequencies above 230 Hz our upper limits exclude compact objects with ellipticities of $10^{-6} {\sqrt{10^{38} \textrm{kg\,m}^2\over I}}$ (corresponding to GW spindowns between $10^{-12}$  Hz/s and $10^{-11}$ Hz/s)  within 100 pc of Earth. Both the ellipticity and the distance ranges span absolutely plausible values and could not have been excluded with other measurements.

We expect the methodology used in this search to serve as a template for the assessment of \EatH run results in the future, for example the next \EatH run, using advanced LIGO data that is being processed as this paper is written. Results of searches for continuous wave signals could also be mined further, probing sub-threshold candidates with a hierarchical series of follow-up searches. This is not the topic of this paper and might be pursued in a forthcoming publication.

\section{Acknowledgments}

The authors gratefully acknowledge the support of the Einstein@Home volunteers, of the United States
National Science Foundation for the construction and operation of the
LIGO Laboratory, the Science and Technology Facilities Council of the
United Kingdom, the Max-Planck-Society, and the State of
Niedersachsen/Germany for support of the construction and operation of
the GEO600 detector, and the Italian Istituto Nazionale di Fisica
Nucleare and the French Centre National de la Recherche Scientifique
for the construction and operation of the Virgo detector. The authors
also gratefully acknowledge the support of the research by these
agencies and by the Australian Research Council, 
the International Science Linkages program of the Commonwealth of Australia,
the Council of Scientific and Industrial Research of India, 
the Istituto Nazionale di Fisica Nucleare of Italy, 
the Spanish Ministerio de Educaci\'on y Ciencia, 
the Conselleria d'Economia Hisenda i Innovaci\'o of the
Govern de les Illes Balears, the Foundation for Fundamental Research
on Matter supported by the Netherlands Organisation for Scientific Research, 
the Polish Ministry of Science and Higher Education, the FOCUS
Programme of Foundation for Polish Science,
the Royal Society, the Scottish Funding Council, the
Scottish Universities Physics Alliance, The National Aeronautics and
Space Administration, the Carnegie Trust, the Leverhulme Trust, the
David and Lucile Packard Foundation, the Research Corporation, and
the Alfred P. Sloan Foundation.

This document has been assigned LIGO Laboratory document number \texttt{LIGO-P1600156-v22}.

\newpage


\newpage
\appendix
\onecolumngrid
\section{Tabular data}
\subsection{Upper limit values}
\label{A:ULs}
\onecolumngrid
\begin{longtable}{|c|c|c@{\hskip 0.1in}|c|c|c@{\hskip 0.1in}|c|c|c@{\hskip 0.1in}|c|c|}
\cline{1-2}\cline{4-5}\cline{7-8}\cline{10-11}
$\boldsymbol{f}$ \textbf{(in Hz)}& $\boldsymbol{h_{0}^{90\%}\times 10^{25}}$ & & $\boldsymbol{f}$ \textbf{(in Hz)} & $\boldsymbol{h_{0}^{90\%}\times 10^{25}}$ & & $\boldsymbol{f}$ \textbf{(in Hz)} & $\boldsymbol{h_{0}^{90\%}\times 10^{25}}$ & & $\boldsymbol{f}$ \textbf{(in Hz)} & $\boldsymbol{h_{0}^{90\%}\times 10^{25}}$\vspace{0.02in}\\
\cline{1-2}\cline{4-5}\cline{7-8}\cline{10-11}
\endhead
\centering
50.063 & 70.3 $\pm$ 12.8 & & 50.563 & 68.4 $\pm$ 12.5 & & 51.063 & 69.3 $\pm$ 12.6 & & 51.563 & 67.5 $\pm$ 12.4 \\
52.063 & 66.9 $\pm$ 12.2 & & 53.063 & 57.6 $\pm$ 10.5 & & 53.563 & 58.9 $\pm$ 10.9 & & 54.063 & 55.3 $\pm$ 10.1 \\
54.563 & 54.0 $\pm$ 9.9 & & 55.063 & 55.7 $\pm$ 10.2 & & 55.563 & 53.3 $\pm$ 9.8 & & 56.063 & 50.9 $\pm$ 9.3 \\
56.563 & 51.8 $\pm$ 9.5 & & 57.063 & 47.5 $\pm$ 8.7 & & 57.563 & 46.9 $\pm$ 8.6 & & 58.063 & 47.1 $\pm$ 8.6 \\
58.563 & 51.5 $\pm$ 9.4 & & 61.063 & 44.8 $\pm$ 8.2 & & 61.563 & 37.4 $\pm$ 6.9 & & 62.063 & 36.5 $\pm$ 6.7 \\
62.563 & 36.0 $\pm$ 6.6 & & 63.063 & 36.3 $\pm$ 6.6 & & 63.563 & 33.8 $\pm$ 6.2 & & 64.063 & 30.6 $\pm$ 5.6 \\
64.563 & 29.8 $\pm$ 5.4 & & 65.063 & 31.5 $\pm$ 5.9 & & 65.563 & 30.8 $\pm$ 5.7 & & 66.063 & 28.3 $\pm$ 5.2 \\
66.563 & 26.5 $\pm$ 4.8 & & 67.063 & 26.5 $\pm$ 4.9 & & 67.563 & 27.3 $\pm$ 5.0 & & 68.063 & 25.7 $\pm$ 4.7 \\
\end{longtable}
\clearpage
\onecolumngrid
\subsection{Cleaned-out frequency bins}
\label{A:cleanedBands}
\twocolumngrid
\begin{longtable}{|l|ccccc|} \hline \hline 
Cause       & $f_{\mathrm{L}}$~(Hz)   &  Harmonics    &  LFS~(Hz)    & HFS~(Hz)           & IFO\\
\hline 
\hline
\endhead
Mains  &  60   &  8   &  1    &  1            & L,H\\
\hline
Wire    & 345   &  1    &  5    &  5            & L\\
Wire   & 346   &  1    &  4    &  4            & H\\
\hline
Electronic & 85.8 & 1 & 0.01 & 0.01 & H\\
Electronic & 89.9 & 1 & 0.06 & 0.06 & H\\
Electronic & 93.29 & 1 & 0.015 & 0.015 & L\\
Electronic & 93.05 & 1 & 0.01 & 0.01 & H\\
Electronic & 93.25 & 1 & 0.01 & 0.01 & H\\
\end{longtable}

\clearpage
\onecolumngrid
\subsection{50mHz signal-frequency bands that did not contribute to results}
\label{A:excluded50mHzBands}
\twocolumngrid
\begin{longtable}{|ll|c|}
\hline
\hline
start & stop & disturbance \\ 
band & band & type \\ 
\hline
\hline
\endhead
50.113	&	&D \\
50.563	&	&D \\
51.013	&	&D \\
51.113	&51.163	&D \\
51.963	&	&D \\
52.413	&52.463	&D \\
52.613	&	&D \\
52.863	&53.113	&D \\
\end{longtable}

\clearpage
\onecolumngrid
\subsection{Signal-frequency ranges where the results might have contributions from fake data}
\label{A:suspectFreqRanges}
\begin{longtable}{|l||cc||cc|cc|cc||c|}
\hline
\hline
line & \multicolumn{2}{c||}{mixed,}   & \multicolumn{2}{c|}{mixed,} & \multicolumn{2}{c|}{all} & \multicolumn{2}{c||}{mixed,} & \\ 
type & \multicolumn{2}{c||}{isolated} & \multicolumn{2}{c|}{left}   & \multicolumn{2}{c|}{fake data} & \multicolumn{2}{c||}{right} & detector\\
\hline
\hline
\endhead
1Hz	& 50.9648	& 51.0352	& 	& 	& 	& 	& 	& 	&L,H	\\
1Hz, 2Hz	& 51.9647	& 52.0353	& 	& 	& 	& 	& 	& 	&L,H	\\
1Hz	& 52.9646	& 53.0354	& 	& 	& 	& 	& 	& 	&L,H	\\
1Hz	& 53.9645	& 54.0355	& 	& 	& 	& 	& 	& 	&L,H	\\
CPU	& 54.4605	& 54.5315	& 	& 	& 	& 	& 	& 	&L,H	\\
1Hz	& 54.9644	& 55.0356	& 	& 	& 	& 	& 	& 	&L,H	\\
D	& 	& 	& 55.715	& 55.785	& 55.785	& 55.815	& 55.815	& 55.885	&H	\\
1Hz	& 55.9643	& 56.0357	& 	& 	& 	& 	& 	& 	&L,H	\\
\end{longtable}

\clearpage

\iftoggle{endauthorlist}{
  %
  %
  \let\author\myauthor
  \let\affiliation\myaffiliation
  \let\maketitle\mymaketitle
  \title{Authors}
  \pacs{}
  
  \newpage
  \maketitle
}

\end{document}